\RequirePackage{etoolbox}
\csdef{input@path}{%
 {sty/}
 {img/}
}%
\csgdef{bibdir}{bib/}

\documentclass[ba]{imsart}
\pubyear{0000}
\volume{00}
\issue{0}
\doi{0000}
\firstpage{1}
\lastpage{1}

\usepackage{enumerate}
\usepackage{xcolor}

\usepackage{booktabs}
\usepackage{multirow}
\usepackage{bm}
\usepackage{natbib}
\usepackage{tikz}
\usetikzlibrary{arrows}
\usetikzlibrary{positioning}
\newdimen\nodeDist{}
\usepackage{amsthm}
\usepackage{amsmath}
\usepackage{natbib}
\usepackage[colorlinks,citecolor=blue,urlcolor=blue,filecolor=blue,backref=page]{hyperref}
\usepackage{amssymb, amsfonts,bbm,bbold}

\usepackage{amsbsy}

\newcommand{\iid}{\stackrel{\mbox{\tiny iid} }{\sim}}

\newcommand{\X}{\mathbf{X}}

\newcommand{\epsy}{\epsilon}

\newcommand{\z}{\mathrm{z}}

\newcommand{\N}{\mbox{{\small\textsc{N}}}}

\newcommand{\E}{\mbox{E}}

\newcommand{\x}{\mathrm{x}}

\newcommand\independent{\protect\mathpalette{\protect\independenT}{\perp}}
\def\independenT#1#2{\mathrel{\rlap{$#1#2$}\mkern2mu{#1#2}}}

\startlocaldefs
\numberwithin{equation}{section}
\theoremstyle{plain}

\endlocaldefs

\begin{document}

\begin{frontmatter}
\title{Bayesian regression tree models for causal inference: regularization, confounding, and heterogeneous effects}
\runtitle{Bayesian regression tree models for causal inference}

\begin{aug}
\author{\fnms{P. Richard} \snm{Hahn}\thanksref{addr1}\ead[label=e1]{prhahn@asu.edu}},
\author{\fnms{Jared S.} \snm{Murray}\thanksref{addr2}\ead[label=e2]{jared.murray@mccombs.utexas.edu}}
\and
\author{\fnms{Carlos M.} \snm{Carvalho}\thanksref{addr2}
\ead[label=e3]{carlos.carvalho@mccombs.utexas.edu}
}

\runauthor{Hahn, Murray and Carvalho}

\address[addr1]{School of Mathematical and Statistical Sciences, Arizona State University
    \printead{e1} 
}

\address[addr2]{McCombs School of Business, University of Texas at Austin
    \printead{e2}
    \printead{e3}
}

\end{aug}

\begin{abstract}

This paper presents a novel nonlinear regression model for estimating heterogeneous treatment effects from observational data, geared specifically towards situations with small effect sizes, heterogeneous effects, and strong confounding. Standard nonlinear regression models, which may work quite well for prediction, have two notable weaknesses when used to estimate heterogeneous treatment effects. First, they can yield badly biased estimates of treatment effects when fit to data with strong confounding. The Bayesian causal forest model  presented in this paper avoids this problem by directly incorporating an estimate of the propensity function in the specification of the response model, implicitly inducing a covariate-dependent prior on the regression function. Second, standard approaches to response surface modeling do not provide adequate control over the strength of regularization over effect heterogeneity. The Bayesian causal forest model permits treatment effect heterogeneity to be regularized separately from the prognostic effect of control variables, making it possible to informatively ``shrink to homogeneity''. We illustrate these benefits via the reanalysis of an observational study assessing the causal effects of smoking on medical expenditures as well as extensive simulation studies.
 \end{abstract}

\begin{keyword}[class=MSC]
\kwd[Primary ]{60K35}
\kwd{60K35}
\kwd[; secondary ]{60K35}
\end{keyword}

\begin{keyword}
\kwd{Bayesian}
\kwd{Causal inference}
\kwd{Heterogeneous treatment effects}
\kwd{Predictor-dependent priors}
\kwd{Machine learning}
\kwd{Regression trees}
\kwd{Regularization}
\kwd{Shrinkage}
\end{keyword}

\end{frontmatter}

\section{Introduction}
The success of modern predictive modeling is founded on the understanding that flexible predictive models must be carefully regularized in order to achieve good out-of-sample performance (low generalization error). In a causal inference setting, regularization is less straightforward, for (at least) two reasons. One, in the presence of confounding, regularized  models originally designed for prediction can bias causal estimates towards some unknown function of high dimensional nuisance parameters \citep{hahn2016regularization}. Two, when the magnitude of response surface variation due to prognostic effects differs markedly from response surface variation due to treatment effect heterogeneity, simple regularization strategies, which may be adequate for good out-of-sample prediction, provide inadequate control of estimator variance for conditional average treatment effects (leading to large estimation error).  


To mitigate these two estimation issues we propose a flexible sum-of-regression-trees --- a {\em forest} --- to model a response variable as a function of a binary treatment indicator and a vector of control variables. To address the first issue, we develop a novel prior for the response surface that depends explicitly on estimates of the propensity score as an important 1-dimensional transformation of the covariates (including the treatment assignment). Incorporating this transformation of the covariates is not strictly necessary in response surface modeling in order to obtain consistent estimators, but we show that it
can substantially improve treatment effect estimation in the presence of moderate to strong confounding, especially when that confounding is driven by targeted selection --- individuals selecting into treatment based on somewhat accurate predictions of the potential outcomes.  

To address the second issue, we represent our regression as a sum of two functions: the first models the {\em prognostic} impact of the control variables (the component of the conditional mean of the response that is unrelated to the treatment effect), while the second represents the treatment effect directly, which itself is a nonlinear function of the observed attributes (capturing possibly heterogeneous effects).  We represent each function as a forest. This approach allows the degree of shrinkage on the treatment effect to be modulated {\em directly} and {\em separately} of the prognostic effect. In particular, under this parametrization, standard regression tree priors shrink towards homogeneous effects. 

In most previous approaches, the prior distribution over treatment effects is induced indirectly, and is therefore difficult to understand and control.  Our approach interpolates between two extremes: Modeling the conditional means of treated and control units entirely separately, or including treatment assignment as ``just another covariate''.
The former precludes any borrowing or regularization entirely, while the second can be rather difficult to understand using flexible models. Parametrizing non- and semiparametric models this way  is attractive regardless of the specific priors in use. 

Comparisons on simulated data show that the new model --- which we call the Bayesian causal forest model --- performs at least as well as existing approaches for estimating heterogenous treatment effects across a range of plausible data generating processes. More importantly, it performs dramatically better in many cases, especially those with strong confounding, targeted selection, and relatively weak treatment effects, which we believe to be common in applied settings. 

In section~\ref{sec:smoking}, we demonstrate how our flexible Bayesian model allows us to make rich inferences on heterogeneous treatment effects, including estimates of average and conditional average treatment effects at various levels, in a re-analysis of data from an observational study of the effect of smoking on medical expenditures.

\subsection{Relationship to previous literature}
As  previously noted, the Bayesian causal forest model directly extends ideas from two earlier papers:  \cite{hill2011bayesian} and \cite{hahn2016regularization}. Specifically, this paper studies the ``regularization-induced confounding'' of \cite{hahn2016regularization} in the context of nonparametric Bayesian models as utilized by \cite{hill2011bayesian}. In terms of implementation, this paper builds explicitly on the work of \cite{chipman2010bart}; see also \cite{gramacy2008bayesian} and \cite{murray2017loglinear}. Other notable work on Bayesian treatment effect estimation includes \cite{gustafson2006curious},\cite{zigler2014uncertainty}, \cite{heckman2014treatment}, \cite{li2014bayesian}, \cite{roy2017bayesian} and  \cite{taddy2016nonparametric}. 

More generally, the intersection between ``machine learning'' and causal inference is a burgeoning research area.  

Papers deploying nonlinear regression (supervised learning) methods in the service of estimation and inference for average treatment effects (ATEs) include targeted maximum likelihood estimation (TMLE) \citep{van2010targetedI, van2010targetedII}, double machine learning \citep{chernozhukov2016double}, and generalized boosting \citep{mccaffrey2004propensity, mccaffrey2013tutorial}. These methods all take as inputs regression estimates of either/or the propensity function or the response surface; in this sense, any advances in predictive modeling have the potential to improve ATE estimation in conjunction with the above approaches. Bayesian causal forests could be used in this capacity as well, although it was designed with conditional average treatment effects in mind. 

More recently, attention has turned to CATE estimation. Notable examples include \cite{taddy2016nonparametric}, who focus on estimating heterogeneous effects from experimental, as opposed to observational data, which is our focus. \cite{su2012facilitating} approach CATE estimation with regression tree ensembles and are in that sense a forerunner of both Bayesian causal forests as well as \cite{wager2018estimation}\footnote{Note that the Bayesian causal forest model is not the Bayesian analogue of the causal random forest method, as both the motivation and fitting process are quite different; both are tree-based methods for estimating CATEs, but the similarities end there. Specifically, \cite{chipman2010bart} is already substantially different than \cite{breiman2001random}, and the ways that BCF modifies BART are simply not analogous to the modifications that causal random forests makes to random forests.},  \cite{athey2019generalized} and \cite{powers2018some}. \cite{wager2018estimation} is notable for providing the first inferential theory for CATE estimators arising from a random forests representation, based on the infinitesimal jackknife \citep{efron2014estimation, wager2014confidence}. \cite{friedberg2018local} extend this approach to locally linear forests. \cite{nie2017quasi} and \cite{kunzel2019metalearners} propose stacking and meta-learning methods, respectively, similar to what TMLE does for the ATE, except tailored to CATE estimation. \cite{shalit2017estimating} develop a neural network-based estimator of CATEs based on a bound of the generalization error in an approach inspired by domain adaptation \citep{ganin2016domain}. \cite{Zaidi2018gaussian} develop a model based on the use of Gaussian processes to directly model the special transformed response (as studied in \cite{athey2019generalized} and \cite{powers2018some}).

The focus of the present paper is to develop a regularization prior for nonlinear models geared specifically towards situations with small effect sizes, heterogeneous effects, and strong confounding. The research above does not focus specifically on this regime, which is an important one in applied settings. 

Finally there are a number of papers that compare and contrast the above methods on real and synthetic data: \cite{wendling2018comparing}, \cite{mcconnell2019estimating}, \cite{aciccomp2016}, \cite{dorie2019automated}, and \cite{ACIC2017} . The results of these studies will be discussed in some detail later, but a general trend is that BART-based methods appear to be a strong default choice for heterogeneous effect modeling.

\section{Problem statement and notation}
Let $Y$ denote a scalar response variable and $Z$ denote a binary treatment indicator variable. Capital Roman letters denote random variables, while realized values appear in lower case, that is, $y$ and $z$. Let $\x$ denote a length $d$ vector of observed control variables. Throughout, we will consider an observed sample of size $n$ independent observations $(Y_i, Z_i, \x_i)$, for $i = 1, \dots n$. When $Y$ or $Z$ (respectively, $y$ or $z$) are without a subscript, they denote length $n$ column vectors; likewise, $\X$ will denote the $n \times d$ matrix of control variables.  

We are interested in estimating various treatment effects. In particular, we are interested in conditional average treatment effects (CATE) --- the amount by which the response $Y_i$ would differ between hypothetical worlds in which the treatment was set to $Z_i = 1$ versus $Z_i = 0$, averaged across subpopulations defined by attributes $\x$. This kind of counterfactual estimand can be formalized in the potential outcomes framework (\cite{imbens2015causal}, chapter 1) by using $Y_i(0)$ and $Y_i(1)$ to denote the outcomes we would have observed if treatment were set to zero or one, respectively.   We make the stable unit treatment value assumption (SUTVA) throughout (excluding interference between units and multiple versions of treatment \citep{imbens2015causal}).  We observe the potential outcome that corresponds to the realized treatment: $Y_i = Z_i Y_i(1) + (1-Z_i) Y_i(0)$.

Throughout the paper we will assume that {\em strong ignorability} holds, which stipulates that
\begin{equation}\label{ignore1}
Y_i(0), Y_i(1) \independent Z_i \mid \X_i.
\end{equation}
and also that 
\begin{equation}\label{ignore2}
0 < \mbox{Pr}(Z_i = 1 \mid \x_i) < 1
\end{equation}
for all $i = 1, \dots, n$. The first condition assumes we have no unmeasured confounders, and the second condition (overlap) is necessary to estimate treatment effects everywhere in covariate space. Provided that these conditions hold, it follows that $\E(Y_i(z) \mid \x_i) = \E(Y_i \mid \x_i, Z_i = z)$ so our estimand may be expressed as 
\begin{equation}\label{TE}
\tau(\x_i) := \E(Y_i \mid \x_i, Z_i = 1) -  \E(Y_i \mid \x_i, Z_i = 0).
\end{equation}

For simplicity, we restrict attention to mean-zero additive error representations
\begin{equation}\label{additive}
Y_i = f(\x_i, Z_i) + \epsy_i,\quad \epsy_i\sim N(0,\sigma^2)
\end{equation}
so that $\E(Y_i \mid \x_i, Z_i=z_i) = f(\x_i, z_i)$. In this context, \eqref{ignore1} can be expressed equivalently as $\epsy_i \independent Z_i \mid \x_i$. 
The treatment effect of setting $z_i = 1$ versus $z_i = 0$ can therefore be expressed as 
$$\tau(\x_i) := f(\x_i, 1) - f(\x_i, 0).$$ 

Our contribution in this paper is a careful study of prior specification for $f$. We propose new prior distributions that improve estimation of the parameter of interest, namely $\tau$. Previous work \citep{hill2011bayesian} advocated using a Bayesian additive regression tree (BART) prior for $f(\x_i, z_i)$ directly. We instead recommend expressing the response surface as
\begin{equation}
\E(Y_i \mid \x_i, Z_i = z_i) = \mu(\x_i, \hat{\pi}(\x_i)) + \tau(\x_i) z_i,
\end{equation}
where the functions $\mu$ and $\tau$ are given independent BART priors and $\hat{\pi}(\x_i)$ is an estimate of the propensity score $\pi(\x_i) = \mbox{Pr}(Z_i = 1 \mid \x_i)$. The following sections motivate this model specification and provide additional context; further modeling details are given in Section \ref{bcf}.

\section{Bayesian additive regression trees for heterogeneous treatment effect estimation}
\cite{hill2011bayesian} observed that under strong ignorability, treatment effect estimation reduces to response surface estimation. That is, provided that a sufficiently rich collection of control variables are available (to ensure strong ignorability), treatment effect estimation can proceed ``merely" by estimating the conditional expectations $\E(Y \mid \x, Z = 1)$ and $\E(Y \mid \x, Z = 0)$. 
Noting its strong performance in prediction tasks, \cite{hill2011bayesian} advocates the use of the Bayesian additive regression tree (BART) model of \cite{chipman2010bart} for estimating these conditional expectations.

BART is particularly well-suited to detecting interactions and discontinuities, can be made invariant to monotone transformations of the covariates, and typically requires little parameter tuning.  \cite{chipman2010bart} provide extensive evidence of BART's excellent predictive performance.  BART has also been used successfully for applications in causal inference, for example \cite{green2012modeling}, \cite{hill2013assessing}, \cite{kern2016assessing}, and \cite{sivaganesan2017subgroup}. It has subsequently been demonstrated to successfully infer heterogeneous and average treatment effects in multiple independent simulation studies \citep{dorie2019automated, wendling2018comparing}, frequently outperforming competitors (and never lagging far behind).

\subsection{Specifying the BART prior}
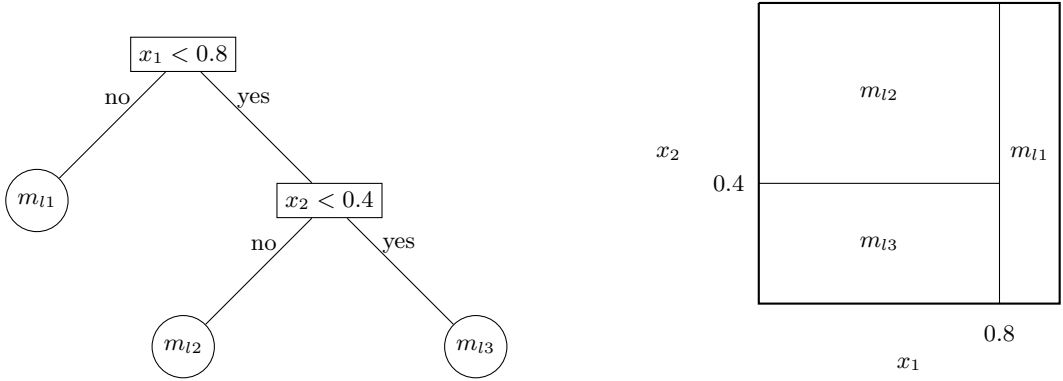
\begin{figure}
\begin{center}
\begin{tikzpicture}[
  scale=0.8,
    node/.style={%
      draw,
      rectangle,
    },
    node2/.style={%
      draw,
      circle,
    },
  ]
    \node [node] (A) {$x_1<0.8$};
    \path (A) ++(-135:\nodeDist) node [node2] (B) {$m_{l1}$};
    \path (A) ++(-45:\nodeDist) node [node] (C) {$x_2<0.4$};
    \path (C) ++(-135:\nodeDist) node [node2] (D) {$m_{l2}$};
    \path (C) ++(-45:\nodeDist) node [node2] (E) {$m_{l3}$};

    \draw (A) -- (B) node [left,pos=0.25] {no}(A);
    \draw (A) -- (C) node [right,pos=0.25] {yes}(A);
    \draw (C) -- (D) node [left,pos=0.25] {no}(A);
    \draw (C) -- (E) node [right,pos=0.25] {yes}(A);
\end{tikzpicture}
\hspace{0.1\linewidth}
\begin{tikzpicture}[scale=3]
\draw [thick, -] (0,1) -- (0,0) -- (1,0) -- (1,1)--(0,1);
\draw [thin, -] (0.8, 1) -- (0.8, 0);
\draw [thin, -] (0.0, 0.4) -- (0.8, 0.4);
\node at (-0.1,0.4) {0.4};
\node at (0.8,-0.1) {0.8};
\node at (0.5,-0.2) {$x_1$};
\node at (-0.3,0.5) {$x_2$};
\node at (0.9,0.5) {$m_{l1}$};
\node at (0.4,0.7) {$m_{l2}$};
\node at (0.4,0.2) {$m_{l3}$};
\end{tikzpicture}
\end{center}
\caption{(Left) An example binary tree, with internal nodes labelled by their splitting rules and terminal nodes labelled with the corresponding parameters $m_{lb}$. (Right) The corresponding partition of the sample space and the step function.}
\label{fig:treestep}
\end{figure}

The BART prior expresses an unknown function $f(\x)$ as a sum of many piecewise constant binary regression trees. (In this section, we suppress $z$ in the notation; implicitly $z$ may be considered as a coordinate of $\x$.)
%
Each tree $T_l,\;1\leq l\leq L$, consists of a set of internal decision nodes which define a partition of the covariate space (say $\mathcal{A}_1,\dots,\mathcal{A}_{B(l)}$), as well as a set of terminal nodes or leaves corresponding to each element of the partition. Further, each element of the partition $\mathcal{A}_b$ is associated a parameter value, $m_{lb}$. Taken together the partition and the leaf parameters define a piecewise constant function: $g_l(x) = m_{lb}\ \text{if}\ x\in \mathcal{A}_b$; see Figure \ref{fig:treestep}. 

Individual regression trees are then additively combined into a single regression {\em forest}: $f(\x)=\sum_{l=1}^L g_l(\x).$ Each of the functions $g_l$ are constrained by their prior to be ``weak learners" in the sense that the prior favors small trees and leaf parameters that are near zero. Each tree follows (independently) the prior described in \cite{chipman1998bayesian}: the probability that a node at depth $h$ splits is given by 
$\eta (1+h)^{-\beta},\;\;\eta \in (0,1),\;\beta\in [0,\infty).$

A variable to split on, as well as a cut-point to split at, are then selected uniformly at random from the available splitting rules.  Large, deep trees are given extremely low prior probability by taking $\eta=0.95$ and $\beta=2$ as in \cite{chipman2010bart}. 
The leaf parameters are assigned independent priors
$m_{lb}\sim \N(0,\sigma^2_m)\;\;\textrm{ where }\sigma_m=\sigma_0/\sqrt{L}$. 
The induced marginal prior for $f(\x)$ is centered at zero and puts approximately 95\% of the prior mass within $\pm 2\sigma_0$ (pointwise), and $\sigma_0$ can be used to calibrate the plausible range of the regression function.  Full details of the BART prior and its implementation are given by \cite{chipman2010bart}.

In our context we are concerned with the impact that the prior over $f(\x, z)$ has on estimating $\tau(\x) = f(\x, 1) - f(\x, 0)$. The choice of BART as a prior over $f$ has particular implications for the induced prior on $\tau$ that are difficult to understand: In particular, the induced prior will vary with the dimension of $\x$ and the degree of dependence with $z$.  In Section \ref{bcf} we propose an alternative parameterization that mitigates this problem. But first, the next section develops a more general framework for investigating the influence of prior specification and regularization on treatment effect estimates.

\section{The central role of the propensity score in regularized causal modeling}\label{ric}
In this section we explore the joint impacts of regularization and confounding on estimation of heterogeneous treatment effects.  We find that including an estimate of the propensity score as a covariate reduces the bias of regularized treatment effect estimates in finite samples.  We recommend including an estimated propensity score as a covariate as routine practice regardless of the particular models or algorithms used to estimate treatment effects since regularization is necessary to estimate heterogeneous treatment effects non- or semiparamaterically or in high dimensions. To illustrate the potential for biased estimation and motivate our fix, we introduce two key concepts: Regularization induced confounding and targeted selection. 
 
\subsection{Regularization-induced confounding}
Since treatment effects may be deduced from the conditional expectation function $f(\x_i, z_i)$, a likelihood perspective suggests that the conditional distribution of $Y$ given $\x$ and $Z$ is sufficient for estimating treatment effects.  While this is true in terms of {\em identification} of treatment effects, the question of estimation with finite samples is more nuanced. In particular, many functions in the support of the prior will yield approximately equivalent likelihood evaluations, but may imply substantially different treatment effects. This is particularly true in a strong confounding-modest treatment effect regime, where the conditional expectation of $Y$ is largely determined by $\x$ rather than $Z$.

Accordingly, the posterior estimate of the treatment effect is apt to be substantially influenced by the prior distribution over $f$ for realistic sample sizes. This issue was explored by \cite{hahn2016regularization} in the narrow context of linear regression with continuous treatment and homogenous treatment effect; they call this phenomenon ``regularization-induced confounding'' (RIC). In the linear regression setting an exact expression for the bias on the treatment effect under standard regularization priors is available in closed form.

\subsection*{Example: RIC in the linear model}
Suppose the treatment effect is homogenous and response and treatment model are both linear:
\begin{equation}\label{linear}
\begin{split}
Y_i &= \tau Z_i + \beta^t \x_i + \varepsilon_i,\\
Z_i &= \gamma^t \x_i + \nu_i;
\end{split}
\end{equation}
where the error terms are mean zero Gaussian and a multivariate Gaussian prior is placed over all regression coefficients. The Bayes estimator under squared error loss is the posterior mean, so we examine the expression for the bias of $\hat{\tau}_{rr} \equiv \E(\tau \mid Y, \z, \X)$. We begin from a standard expression for the bias of the ridge estimator, as given, for example, in \cite{giles1979mean}. Write $\theta = (\tau, \beta^t)^t$, $\tilde{\mathbf{X}} = \begin{pmatrix} \z & \mathbf{X} \end{pmatrix}$ and let $\theta \sim \N(0, \mathbf{M}^{-1})$. Then the bias of the Bayes estimator is
\begin{equation}
\mbox{bias}(\hat{\theta}_{rr}) =  -(\mathbf{M} + \tilde{\mathbf{X}}^t\tilde{\mathbf{X}})^{-1}\mathbf{M}\theta\label{eq:ridgebias}
\end{equation}
where the bias expectation is taken over $Y$, conditional on $\X$ and all model parameters.

Consider  $M = \begin{pmatrix} 0 & 0\\ 0 & \mbox{I}_p\end{pmatrix}$, where $\mbox{I}_p$ denotes a $p$-by-$p$ identity matrix, which corresponds to a ridge prior (with ridge parameter $\lambda = 1$ for simplicity) on the control variables and a non-informative ``flat'' prior over the first element ($\tau$, the treatment effect). Plugging this into the bias equation \eqref{eq:ridgebias} and noting that
 $$(\mathbf{M} + \tilde{\mathbf{X}}^t\tilde{\mathbf{X}})^{-1} = \begin{pmatrix} \z^t\z & \z^t\X \\ \X^t\z & \X^t\X + \mathbf{I}_p \end{pmatrix}^{-1} $$ 
 we obtain
\begin{equation}
\mbox{bias}(\hat{\tau}_{rr}) = -\left((\z^t\z)^{-1}\z^t\mathbf{X}\right)(\mathbf{I} + \mathbf{X}^t(\mathbf{X} - \hat{\mathbf{X}}_{\z}))^{-1}\beta,
\end{equation}
where $\hat{\mathbf{X}}_{\z} = \z(\z^t\z)^{-1}\z^t\mathbf{X}$. Notice that the leading term $\left((\z^t\z)^{-1}\z^t\mathbf{X}\right)$ is a vector of regression coefficients from $p$ univariate regressions predicting $X_j$ given $z$. With completely randomized treatment assignment these terms will tend to be near zero (and precisely zero in expectation over $Z$). This ensures that the ridge estimate of $\tau$ is nearly unbiased, despite the fact that the middle matrix is generally nonzero. However, in the presence of selection, some of these regression coefficients will be non-zero due to the correlation between $Z$ and the covariates in $\X$. As a result, the bias of $\hat{\tau}_{rr}$ will depend on the form of the design matrix and unknown nuisance parameters $\beta$.

The problem here is not simply that $\hat\tau_{rr}$ is biased --- after all, the insight behind regularization is that some bias can actually improve our average estimation error. Rather, the problem is that the degree of bias is not under the analyst's control (as it depends on unknown nuisance parameters). The use of a naive regularization prior in the presence of counfounding can unwittingly induce extreme bias in estimation of the target parameter, {\em even when all the confounders are measured and the parametric model is correctly specified}.

In more complicated nonparametric regression models with heterogeneous treatment effects a closed-form expression of the bias is not generally available; see \cite{yang2015semiparametric} and \cite{chernozhukov2016double} for related results in a partially linear model where effects are homogenous but the $\beta^t\x$ term above is replaced by a nonlinear function. However, note that both of these theoretical results consider {\em asymptotic} bias in semi- and non-parametric Bayesian and frequentist inference; our attention here to the simple case of the linear model shows that the phenomenon occurs in finite samples even in a parametric model. That said, the RIC phenomenon can be reliably recreated in nonlinear, semiparametric settings. The easiest way to demonstrate this is by considering scenarios where selection into treatment is based on expected outcomes under no treatment, a situation we call {\em targeted selection}.

\subsection{Targeted selection}\label{sec:targetedselection}

Targeted selection refers to settings where treatment is assigned based on a prediction of the outcome in the {\em absence} of treatment, given measured covariates. That is, targeted selection asserts that treatment is being assigned, in part, based on an estimate of the expected potential outcome $\mu(\x) := \E(Y(0) \mid \x)$ and that the probability of treatment is generally increasing or decreasing as a function of this estimate. 
We suspect this selection process is quite common in practice; for example, in medical contexts where risk factors for adverse outcomes are well-understood physicians are more likely to assign treatment to patients with worse expected outcomes in its absence.

\begin{figure}
\includegraphics[width=2in]{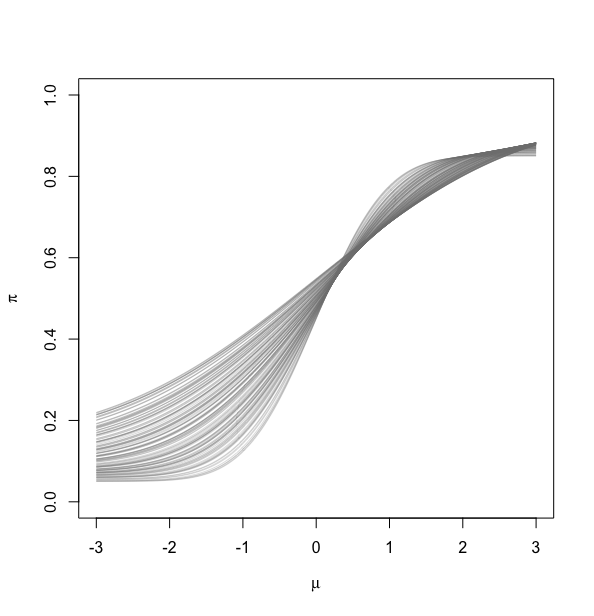}
\caption{For any value of $\tilde{\x}$, the propensity score $\pi(\mu, \tilde{\x})$ is monotone in the prognostic function $\mu$. Here, many realizations of this function are shown for different values of $\tilde{\x}$.} \label{targeted}
\end{figure}

Targeted selection implies that there is a particular functional relationship between the propensity score $\pi$ and the expected outcomes without treatment $\mu$. In particular, suppose for simplicity that there exists a change of variables $\x \rightarrow (\mu(\x), \tilde{\x})$ that takes the prognostic function $\mu(\x)$ to the first element of the covariate vector.  Then targeted selection says  that for every $\tilde{\x}$, the propensity function $\E(Z \mid \x) = \pi(\mu, \tilde{\x})$ is (approximately) monotone in $\mu$; see Figure \ref{targeted} for a visual depiction. If the relationship is strictly monotone so that $\pi$ is invertible in $\mu$ for any $\tilde x$, this in turn implies that $\mu(\x)$ is a function of $\pi(\x)$. 


\begin{figure}
\includegraphics[width=2in]{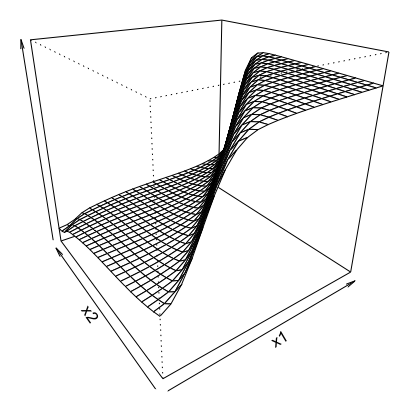}
\includegraphics[width=2in]{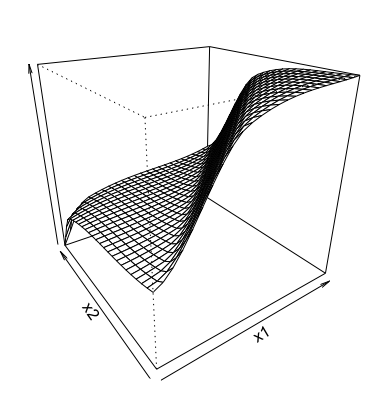}
\caption{Left panel: The propensity function, $\pi$, shown for various values of $\tilde{\x}$. The ``shelf'' at the line $x_1 = x_2$ is a complex shape for many regression methods to represent. Right panel: the analogous plot for the prognostic function $\mu$. Note the similar shapes due to targeted selection; the $\pi$ function falls between $0$ to $1$, while the $\mu$ function ranges from $-3$ to $3$.} \label{persp}
\end{figure}

\subsubsection{Targeted selection and RIC in the linear model}
To help understand how targeted selection leads to RIC, it is helpful to again consider the linear model. There, one can describe RIC in terms of three components: the coefficients defining the propensity function $\E(Z\mid \x) = \gamma\x$, the coefficients defining the prognostic function, $\E(Y\mid Z=0, \x=x)$, and the strength of the selection as measured by $\mbox{Var}(Z\mid \x) = \mbox{Var}(\nu)$. Specifically, note the identity
\begin{equation}\label{linear.bias}
\E(Y \mid \x, Z) = (\tau + b)Z + (\beta - b\gamma)^t\x - b(Z - \gamma^t\x) = \hat{\tau}Z + \hat{\beta}^t\x - \hat{\epsilon},
\end{equation}
which is true for any value of the scalar parameter $b$, the bias of $\hat{\tau}$. Intuitively, if neighborhoods of $\hat{\beta} = (\beta - b\gamma)$ have higher prior probability than $\beta$ and $\mbox{Var}(\hat{\epsilon}) = b^2 \mbox{Var}(\nu)$ is small on average relative to $\sigma^2$, then the posterior distribution for $\tau$ is apt to be biased toward $\hat \tau = \tau+b$.

The bias will be large precisely when confounding is strong and the selection is targeted: For non-negligible bias the term $b^2 \mbox{Var}(\nu)$ is smallest when $\mbox{Var}(\nu)$ is small, that is, when selection (hence, confounding) is strong. For priors on $\beta$ that are centered at zero ---which is overwhelmingly the default --- the $(\beta - b\gamma)$ term can be made most favorable with respect to the prior when the vector $\beta$ and $\gamma$ have the same direction, which corresponds to perfectly targeted selection. 

\subsubsection{Targeted selection and RIC in nonlinear models} To investigate RIC in more complex regression settings, we start with a simple 2-d example characterized by targeted selection:
%
%
\subsubsection*{Example 1: $d = 2$, $n = 250$, homogeneous effects}
Consider the following simple data generating process:
\small
\begin{equation}
\begin{split}
Y_i &= \mu(x_1,x_2) - \tau Z_i + \epsilon_i, \\
\E(Y_i \mid x_{i1}, x_{i2}, Z_i = 1) & = \mu(x_1, x_2),\\
\E(Z_i  \mid x_{i1}, x_{i2}) &= \pi(\mu(x_{i1}, x_{i2}), x_{i1}, x_{i2}),\label{eq:ricex}\\
&=0.8\Phi\left(\frac{\mu(x_{i1}, x_{i2})}{0.1(2-x_{i1}-x_{i2})+0.25}\right) + 0.025(x_{i1} + x_{i2}) + 0.05\\
\epsilon_i &\iid \N(0, 1), \; \;x_{i1}, x_{i2} \iid \mbox{Uniform}(0, 1).\\
\end{split}
\end{equation}
\normalsize
Suppose that in \eqref{eq:ricex} $Y$ is a continuous biometric measure of heart distress, $Z$ is an indicator for having received a heart medication, and $x_1$ and $x_2$ are systolic and diastolic blood pressure (in standardized units), respectively. Suppose that it is known that the {\em difference} between these two measurements is prognostic of high distress levels, with positive levels of $x_1 - x_2$ being a critical threshold. At the same time, suppose that prescribers are targeting the drug towards patients with high levels of diagnostic markers, so the probability of receiving the drug is an increasing function in $\mu$. Figure \ref{persp} shows $\pi$ as a function of $x_1$ and $x_2$; figure \ref{targeted} shows the relationship between $\mu$ and $\pi$ for various values of $\tilde{\x} = x_1 + x_2$.


We simulated 200 datasets of size $n= 250$ according to this data generating process with $\tau = -1$. With only a few covariates, low noise, and a relatively large sample size, we might expect most methods to perform well here. 
Table~\ref{table.ex1} shows that standard, unmodified BART exhibits high bias and root mean squared error (RMSE) as well as poor coverage of 95\% credible intervals. Our proposed fix (detailed below) improves on both estimation error and coverage, primarily by including an estimate of $\pi$ as a covariate.


\begin{table}
\caption{The standard BART prior exhibits substantial bias in estimating the treatment effect, poor coverage of 95\% posterior (quantile-based) credible intervals, and high root mean squared error (rmse). A modified BART prior (denoted BCF) allows splits in an estimated propensity score; it performs markedly better on all three metrics.}\label{table.ex1}
\begin{tabular}{lcrc}
Prior &bias& coverage & rmse\\
\hline
BART &0.27&  65\% & 0.31\\
BCF&  0.14&95\%&0.21
\end{tabular}
\end{table}

\begin{figure}
\includegraphics[width=2in]{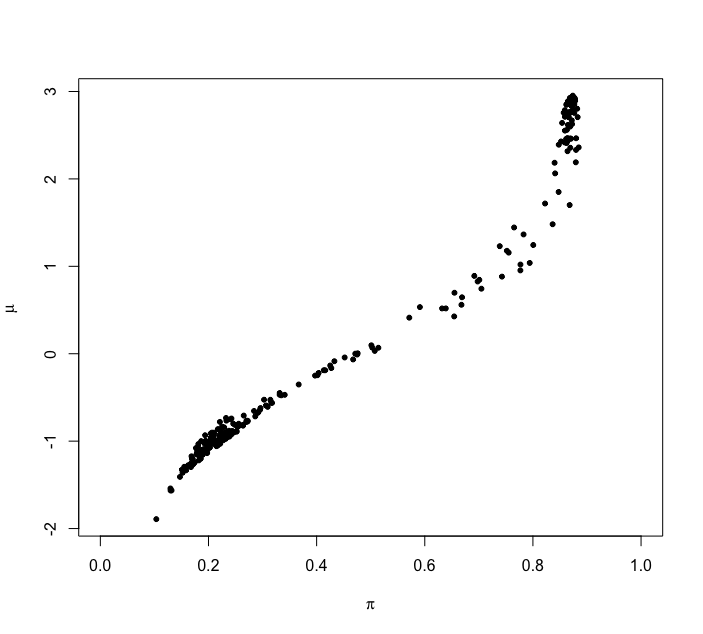}
\caption{This scatterplot depicts $\mu(\x) = \E(Y \mid Z = 0, \x)$ and $\pi(\x) = \E(Z \mid \x)$ for a realization from the data generating process from the above example. It shows clear evidence of targeted selection. Such plots, based on estimates $(\hat{\mu}, \hat{\pi})$ can provide evidence of (strong) targeted selection in empirical data.}\label{scatter} 
\end{figure}

\begin{figure}
\includegraphics[width=2in]{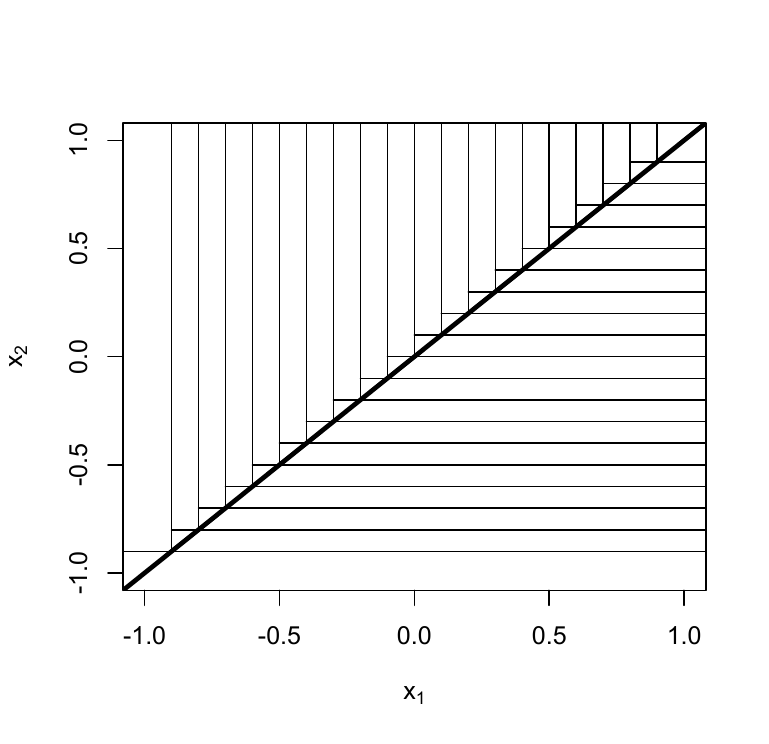}
\caption{Many axis-aligned splits are required to approximate a step function (or near-step function) along the diagonal in the outcome model, as in Fig.~\ref{persp} (right panel). Since these two regions correspond also to disparate rates of treatment, tree-based regularized regression is apt to overstate the treatment effect.}\label{figure.treepart}
\end{figure}

What explains BART's relatively poor performance on this DGP? First, strong confounding and targeted selection implies that $\mu$ is approximately a monotone function of $\pi$ alone (Figure~\ref{scatter}). However, $\pi$ (and hence $\mu$) is difficult to learn via regression trees --- it takes many axis-aligned splits to approximate the ``shelf'' across the diagonal  (see Figure \ref{figure.treepart}), and the BART prior specifically penalizes this kind of complexity. At the same time, due to the strong confounding in this example a single split in $Z$ can stand in for many splits on $x_1$ and $x_2$ that would be required to approximate $\mu(\x)$. These simpler structures are favored by the BART prior, leading to RIC.




Before discussing how we reduce RIC, we note that this example is somewhat stylized in that we designed it specifically to be difficult to learn for tree-based models.  Other models might suffer less from RIC on this particular example. However, any informative, sparse, or nonparametric prior distribution -- any method that imposes meaningful regularization -- is susceptible to similar effects, as they prioritize some data-generating processes at the expense of others.  Absent prior knowledge of the form of the treatment assignment and outcome models, it is impossible to know {\em a prior} whether RIC will be an issue. Fortunately it is  straightforward to minimize the risk of bias due to RIC.

\subsection{Mitigating RIC with covariate-dependent priors}
Finally, we arrive at the role of the propensity score in a regularized regression context. 
The potential for RIC is strongest when $\mu(\x)$ is exactly or approximately a function of $\pi(\x)$ and when the composition of the two has relatively low prior support.  This can lead the model to misattribute the variability of $\mu$, in the direction of $\pi$, to $Z$. A natural solution to this problem would be to include $\pi(\x)$ as a covariate,
so that it is penalized equitably with changes in the treatment variable $Z$. That is, when evaluating candidate functions for our estimate of $\E(Y \mid \x, z)$ we want representations involving $\pi(\x)$ to be regularized/penalized the same as representations involving $z$. Of course $\pi$ is unknown and must be estimated, but this is a straightforward regression problem. Note also that so-called ``unlabeled'' data can be brought to bear here, meaning $\pi$ can be estimated from samples of $(Z, X)$ for which the $Y$ value is unobserved, provided the sample is believed to arise from the relevant population.

\subsubsection{Mitigating RIC in the linear model}
Given an estimate of the propensity function $\hat{z}_i \approx \gamma^t \x_i $, we consider the over-complete regression that includes as regressors both $z$ and $\hat{\z}$. Our design matrix becomes
$$\tilde{\mathbf{X}} = \begin{pmatrix} \z &\hat{\z} & \mathbf{X} \end{pmatrix}.$$ 
This covariate matrix is degenerate because $\hat{z}$ is in the column span of $\mathbf{X}$ by construction. In a regularized regression problem this degeneracy is no obstacle. Applying the expression for the bias from above, with a flat prior over the coefficient associated with $\hat{\z}$, yields 
\begin{equation*}
\mbox{bias}(\hat{\tau}_{rr}) = -\left \lbrace (\tilde{\z}^t\tilde{\z})^{-1}\tilde{\z}^t\mathbf{X}\right \rbrace_1(\mathbf{I} + \mathbf{X}^t(\mathbf{X} - \hat{\X}_{\z}))^{-1}\beta = 0,
\end{equation*}
where $\tilde{\z} = (\z \;\; \hat{\z})$ and $\left \lbrace (\tilde{\z}^t\tilde{\z})^{-1}\tilde{\z}^t\mathbf{X}\right \rbrace_1$ denotes the top row of $\left \lbrace (\tilde{\z}^t\tilde{\z})^{-1}\tilde{\z}^t\mathbf{X}\right \rbrace$, which corresponds to the regression coefficient associated with $z$ in the two variable regression predicting $X_j$ given $\tilde{z}$.  Because $\hat{z}$ captures the observed association between $z$ and $\x$, $z$ is conditionally independent of $\x$ given $\hat{z}$, from which we conclude that these regression coefficients will be zero. See \cite{yang2015semiparametric} for a similar de-biasing strategy in a partially linear semiparametric context. 

\subsubsection{Mitigating RIC in nonlinear models}

The same strategy also proves effective in the nonlinear setting --- simply by including an estimate of the propensity score as a covariate in the BART model, the RIC effect is dramatically mitigated, as can be seen in the second row of Table \ref{table.ex1}. From a Bayesian perspective, this is simply a judicious variable transformation since our regression model is specified conditional on both $Z$ and $\x$ --- we are not obliged to consider uncertainty in our estimate of $\pi$ to obtain valid posterior inference.  We obtain another example of a covariate dependent prior, similar to Zellner's $g$-prior (albeit motivated by very different considerations). 
%
See section \ref{discussion} for additional discussion of this point. Finally, we believe that including an estimated propensity score will be cheap insurance against RIC when estimating treatment effects using outcome models under other nonparametric priors and using more general nonparametric/machine learning approaches. 

To summarize, although it has long been known that the propensity score is a sufficient dimension reduction for estimation of the ATE -- and that combining estimates of the response surface and propensity score can improve estimation of average treatment effects \citep{bang2005doubly}, we find that incorporating an estimate of the propensity score into estimation of the response surface can improve estimation of average treatment effects in finite samples.  As we will demonstrate in Section~\ref{sec:empirical}, these benefits also accrue when estimating (heterogeneous) conditional average treatment effects.  Estimating heterogenous effects also calls for careful consideration of regularization applied to the treatment effect function, which we consider in the next section.

\section{Regularization for heterogeneous treatment effects: Bayesian causal forests}\label{bcf}

In much the same way that a direct BART prior on $f$ does not allow careful handling of confounding, it also does not allow separate control over the discovery of heterogeneous effects because there is no explicit control over how $f$ varies in $Z$. Our solution to this problem is a simple re-parametrization that avoids the indirect specification of the prior over the treatment effects:
\begin{equation}
f(\x_i, z_i) = \mu(\x_i) + \tau(\x_i) z_i.\label{eq:directpar}
\end{equation}
This model can be thought of as a linear regression in $z$ with covariate-dependent functions for both the slope and the intercept. Writing the model this way sacrifices nothing in terms of expressiveness, but permits independent priors to be placed on $\tau$, which is precisely the treatment effect:
\begin{equation}
\E(Y_i \mid \x_i, Z_i = 1) -  \E(Y_i \mid \x_i, Z_i = 0) = \lbrace \mu(\x_i) + \tau(\x_i)\rbrace - \mu(\x_i) = \tau(\x_i).
\end{equation}
Under this model, $\mu(\x) = E(Y\mid Z=0, X=x)$ is a prognostic score in the sense of \cite{hansen2008prognostic}, another interpretable quantity, to which we apply a prior distribution independent of $\tau$ (as detailed below).
Based on the observations of the previous section, we further propose specifying the model as 
\begin{equation}
f(\x_i, z_i) = \mu(\x_i, \hat{\pi}_i) + \tau(\x_i) z_i,\label{eq:bcfmodel}
\end{equation}
where $\hat{\pi}_i$ is an estimate of the propensity score. 

While we will use variants of BART priors for $\mu$ and $\tau$ (see section \ref{sec:bcfprior}), this parameterization has many advantages in general, regardless of the specific priors.  The most obvious advantage is that the treatment effect is an explicit parameter of the model, $\tau(\x)$, and as a result we can specify an appropriate prior on it directly. A similar idea has been proposed previously for non-tree based models in \cite{imai2013estimating}. Before turning to the details of our model specification, we first contrast this parameterization with two common alternatives. 

\subsection{Parameterizing regression models of heterogeneous effects}
There are two common modeling strategies for estimating heterogeneous effects. The first we discussed above: treat $z$ as ``just another covariate'' and specify a prior on $f(\x_i, z_i)$, e.g. as in \cite{hill2011bayesian}.  The second is to fit entirely separate models to the treatment and control data: $(Y\mid Z=z, \x)\sim N(f_z(\x_i), \sigma^2_z)$ with independent priors over the parameters in the $z=0$ and $z=1$ models. In this section we argue that neither approach is satisfactory and propose  the model in \eqref{eq:bcfmodel} as a reasonable interpolation between the two.  (See \cite{kunzel2019metalearners} for a related discussion comparing these two approaches in a non-model-based setting.)

It is instructive to consider (\ref{eq:directpar}) as a nonlinear regression analogue of the common strategy of parametrizing contrasts (differences) and aggregates (sums) rather than group-specific location parameters. 
Specifically, consider a two-group difference-in-means problem: 
\begin{equation}
\begin{split}
Y_{i1} &\iid \N(\mu_1, \sigma^2)\\
Y_{j2} &\iid \N(\mu_2, \sigma^2).
\end{split}
\end{equation}
Although the above parameterization is intuitive, if the estimand of interest is $\mu_1 - \mu_2$, the implied prior over this quantity has variance strictly greater than the variances over $\mu_1$ or $\mu_2$ individually. This is plainly nonsensical if the analyst has no subject matter knowledge regarding the individual levels of the groups, but has strong prior knowledge that $\mu_1 \approx \mu_2$. This is common in a causal inference setting: If the data come from a randomized experiment where $Y_1$ constitutes a control sample and $Y_2$ a treated sample, then subject matter considerations will typically limit the plausible range of treatment effects $\mu_1-\mu_2$.

The appropriate way to incorporate that kind of knowledge is simply to reparametrize:
\begin{equation}
\begin{split}
Y_{i1} &\iid \N(\mu + \tau, \sigma^2)\\
Y_{j2} &\iid \N(\mu, \sigma^2)
\end{split}
\end{equation}
whereupon the estimand of interest becomes $\tau$, which can be given an informative prior centered at zero with an appropriate variance. Meanwhile, $\mu$ can be given a very vague (perhaps even improper) prior. 

While the nonlinear modeling context is more complex, the considerations are the same: our goal is simultaneously to let $\mu(\x)$ be flexibly learned (to adequately deconfound and obtain more precise inference), while appropriately regularizing $\tau(\x)$, which we expect, a priori, to be relatively small in magnitude and ``simple'' (minimal heterogeneity). Neither of the two more common parametrizations permit this: Independent estimation of $f_0(\x)$ and $f_1(\x)$ implies a highly vague prior on $\tau(\x) = f_1(\x) - f_0(\x)$; i.e. a Gaussian process prior on each would imply a twice-as-variable Gaussian process prior on the difference, as in the simple example above. Estimation based on the single response surface $f(\x, z)$ often does not allow direct point-wise control of $\tau(\x) = f(\x, 1) - f(\x, 0)$ at all.  In particular, with a BART prior on $f$ the induced prior on $\tau$ depends on incidental features such as the size and distribution of the covariate vector $\x$.   

\subsection{{Prior specfication}}\label{sec:bcfprior}


With the model parameterized as in \eqref{eq:bcfmodel}, we can specify different BART priors on $\mu$ and $\tau$.  For $\mu$ we use the default suggestions in \citep{chipman2010bart} (200 trees, $\beta=2$, $\eta = 0.95$), except that we place a half-Cauchy prior over the scale of the leaf parameters with prior median equal to twice the marginal standard deviation of $Y$ \citep{gelman2006prior,polson2012half}. We find that inference over $\tau$ is typically insensitive to reasonable deviations from these settings, so long as the prior is not so strong that deconfounding does not take place.

For $\tau$, we prefer stronger regularization. First, we use fewer trees (50 versus 200), as we generally believe that patterns of treatment effect heterogeneity are relatively simple.  Second, we set the depth penalty $\beta=3$ and splitting probability $\eta = 0.25$ (instead of $\beta=2$ and $\eta=0.95$) to shrink more strongly toward homogenous effects (the extreme case where none of the trees split at all corresponds to purely homogenous effects). Finally, we replace the half-Cauchy prior over the scale of $\tau$ with a half Normal prior, pegging the prior median to the marginal standard deviation of $Y$. 
(In the absence of prior information about the plausible range of treatment effects we expect this to be a reasonable uppper bound.) 

\subsection{Data-adaptive coding of treatment assignment}

A less-desirable feature of the model in Eq.~\eqref{eq:bcfmodel} is that different inferences on $\tau(\x)$ can obtain if we code the treatment indicator as zero for treated cases and one for controls, or as $\pm 1/2$ for treated/control units, or any of the other myriad specifications for $Z_i$ that still result in $\tau(\x)$ being the treatment effect. When there is a clear reference treatment level one might think of this as a feature, not a bug, but this is often not the case (such as when comparing two active treatments). Because $\mu $ and $\tau$ alias one another, as under targeted selection, the choice of treatment coding can be meaningfully impact posterior inferences, especially when the treated and control response variables have dramatically different marginal variances. 

Fortunately, an invariant parameterization is possible, which treats the coding of $Z$ as a variable to be estimated:
\begin{gather}
\begin{split}
y_i = \mu(\x_i) + \tilde\tau(\x_i) b_{z_i} + \epsilon_i,\quad \epsilon_i\sim N(0, \sigma^2)\\
b_0\sim N(0, 1/2),\quad b_1\sim N(0, 1/2)\label{eq:pxmodel}
\end{split}
\end{gather}
The treatment effect function in this parameter expanded model is 
\[
\tau(\x_i) = (b_1 - b_0)\tilde\tau(\x_i).
\]
Noting that $b_1-b_0\sim N(0, 1)$ we still obtain a half Normal prior distribution for the scale of the leaf parameters in $\tau$ as in the previous subsection, and we can adjust the scale of the half normal prior (e.g. to fix the scale at one marginal standard deviation of $Y$ as above) using a fixed scale in the leaf prior for $\tilde\tau$. Posterior inference in this model requires only minor adjustments to \cite{chipman2010bart}'s Bayesian backfitting MCMC algorithm.  Specifically, note that conditional on $\tau$, $\mu$ and $\sigma$, updates for $b_0$ and $b_1$ follow from standard linear regression updates, with a two-column design matrix with columns $\left (\tau(\x_i) z_i, \; \tau(\x_i) (1-z_i) \right)$ (no intercept) and the ``residual'' $y_i - \mu(\x_i)$ acting as the response variable.

Our experiments below all use this parameterization, and it is the default implmentation in our software package.

\section{Empirical evaluations}\label{sec:empirical}
In this section, we provide a more extensive look at how BCF compares to various alternatives. In Section~\ref{sec:oursim} we compare BCF, generalized random forests \citep{athey2019generalized}, and a linear model with all three-way interactions as plausible methods for estimating heterogeneous treatment effects with measures of uncertainty. We also consider three specifications of BART:  the standard response surface BART that considers the treatment variable as ``just another covariate'', one where separate BART models are fit to the treatment and control arms of the data, and one where an estimate of the propensity score is included as a predictor. In Section~\ref{sec:contest} we report on the results of two separate data analysis challenges, where the entire community was invited to submit methods for evaluation on larger synthetic datasets with heterogeneous treatment effects.  In both simulation settings we find that BCF performs well under a wide range of scenarios.

In all cases the estimands of interest are either conditional average treatment effects  for individual $i$ accounting for all the variables, estimated by the posterior mean  treatment effect $\hat\tau(\x_i)$, or sample subgroup average treatment effects estimated by
$
\sum_{i\in \mathcal{S}} \hat\tau(\x_i),
$
where $\mathcal{S}$ is the subgroup of interest. Credible intervals are computed from MCMC output.

\subsection{Simulation studies}\label{sec:oursim}
We evaluated three variants of BART, the causal random forest model of \cite{athey2019generalized} (using the default specification in the \texttt{grf} package), and a regularized linear regression with up to three way interactions. We consider eight distinct, but closely related, data generating processes, corresponding to the various combinations of toggling three two-level settings: homogeneous versus heterogeneous treatment effects, a linear versus nonlinear conditional expectation function, and two different sample sizes ($n = 250$ and $n = 500$). Five variables comprise $\x$; the first three are continuous, drawn as standard normal random variables, the fourth is a dichotomous variable and the fifth is unordered categorical, taking three levels (denoted 1,2,3). The treatment effect is either
  \[
   \tau(\x) = \left\{\begin{array}{ll}
        3, & \text{homogeneous} \\
        1 + 2 x_2x_5, & \text{heterogeneous,}
        \end{array} \right .
        \] 
  the prognostic function is either
  \[
   \mu(\x) = \left\{\begin{array}{ll}
        1 + g(x_4) + x_1x_3, & \text{linear} \\
        -6 + g(x_4) + 6 |x_3 -1|, & \text{nonlinear,}
        \end{array} \right .
        \] 
where $g(1) = 2$, $g(2) = -1$ and $g(3) = -4$, and the propensity function is
\[
\pi(\x_i) = 0.8 \Phi(3\mu(\x_i)/s - 0.5x_1) + 0.05 + u_i/10
\]
where $s$ is the standard deviation of $\mu$ taken over the observed sample and $u_i \sim \mbox{Uniform}(0,1)$.

To evaluate each method we consider three criteria, applied to two different estimands. First, we consider how each method does at estimating the (sample) average treatment effect (ATE) according to root mean square error, coverage, and average interval length. Then, we consider the same criteria, except applied to estimates of the conditional average treatment effect (CATE), averaged over the sample. Results are based on 200 independent replications for each DGP. Results are reported in Tables \ref{sim.linear} (for the linear DGP) and \ref{sim.nonlinear} (for the nonlinear DGP). The important trends are as follows:
\begin{itemize}
\item BCF or ps-BART benefit dramatically by explicitly protecting against RIC;
\item BART-$(f_0,f_1)$ and causal random forests both exhibit subpar performance;
\item all methods improve with a larger sample;
\item BCF priors are extra helpful at smaller sample sizes, when estimation is difficult;
\item the linear model dominates when correct, but fares extremely poorly when wrong;
\item BCF's improvements over ps-BART are more pronounced in the nonlinear DGP;
\item BCF's average interval length is notably smaller than the ps-BART interval, usually (but not always) with comparable coverage. 
\end{itemize}

\begin{table}[h!]
\footnotesize\addtolength{\tabcolsep}{-5pt}
\caption{Simulation study results when the true DGP is a linear model with third order interactions. Root mean square estimation error (rmse), coverage (cover) and average interval length (len) are reported for both the average treatment effect (ATE) estimates and the conditional average treatment effect estimates (CATE).}\label{sim.linear}
\begin{tabular*}{\linewidth}{ @{\extracolsep{\fill}} ll *{13}c @{}}

 &  & \multicolumn{6}{c}{Homogeneous effect} &
\multicolumn{6}{c}{Heterogeneous effects}  \\
\toprule
$n$ & Method & \multicolumn{3}{c}{ATE} &
\multicolumn{3}{c}{CATE} & \multicolumn{3}{c}{ATE} & \multicolumn{3}{c}{CATE} \\
\midrule \midrule 
 & &rmse& cover&len&rmse&cover&len&rmse&cover&len&rmse&cover&len\\
\cmidrule{3-6} \cmidrule{7-10} \cmidrule{11-14} \cmidrule{15-15}
\multirow{4}{*}{$250$} & BCF   & 0.21    &    0.92 &0.91 &       0.48  &       0.96& 2.0&  0.27    &    0.84 &0.99    &    1.09       &  0.91 & 3.3 \\
                         & ps-BART &0.22    &    0.94 &0.97  &      0.44  &       0.99& 2.3&  0.31 &       0.90 &1.13  &      1.30  &       0.89 &3.5\\
                         & BART &0.34   &     0.73 &0.94  &      0.54  &       0.95 &2.3 &0.45  &      0.65 &1.10  &      1.36  &       0.87 & 3.4\\
                         & BART $(f_0,f_1)$ &0.56    &    0.41 &0.99  &      0.92 &        0.93 &3.4& 0.61 &       0.44& 1.14  &      1.47  &       0.90 &4.5\\
                         & Causal RF   &0.34   &     0.73& 0.98  &      0.47  &       0.84& 1.3& 0.49 &       0.68& 1.25 &       1.58  &       0.68 &2.4\\
                         & LM + HS  &0.14   &     0.96& 0.83 &       0.26   &      0.99& 1.7 & 0.17 &       0.94& 0.89 &       0.33 &        0.99& 1.9\\
                         \midrule 
\multirow{4}{*}{$500$}    & BCF&    0.16    &    0.88& 0.60&        0.38  &       0.95& 1.4 & 0.16 &       0.90 &0.64  &      0.79  &       0.89 &2.4\\
                         & ps-BART &0.18  &      0.86& 0.63 &       0.35  &       0.99 &1.8& 0.16 &       0.90 &0.69   &     0.86 &        0.95 &2.8\\
                         & BART &0.27       & 0.61& 0.61      &  0.42        & 0.95 &1.8& 0.25       & 0.76 &0.67     &   0.88       &  0.94 &2.8\\
                         & BART $(f_0,f_1)$ &0.47     &   0.21& 0.66  &      0.80  &       0.93 &3.1 &0.42  &      0.42& 0.75  &     1.16   &      0.92 &3.9\\
                         & Causal RF   &0.36  &      0.47& 0.69   &     0.52  &       0.75& 1.2 &0.40 &       0.59 &0.88  &      1.30  &       0.71 &2.1\\
                         & LM + HS  &0.11     &   0.96 &0.54     &   0.18       &  0.99& 1.0 & 0.12    &    0.93 &0.59      &  0.22       &  0.98 &1.2\\
\bottomrule
\end{tabular*}

\end{table}

\begin{table}[h!]
\footnotesize\addtolength{\tabcolsep}{-5pt}
\caption{Simulation study results when the true DGP is nonlinear. Root mean square estimation error (rmse), coverage (cover) and average interval length (len) are reported for both the average treatment effect (ATE) estimates and the conditional average treatment effect estimates (CATE).}\label{sim.nonlinear}
\begin{tabular*}{\linewidth}{ @{\extracolsep{\fill}} ll *{13}c @{}}

 &  & \multicolumn{6}{c}{Homogeneous effect} &
\multicolumn{6}{c}{Heterogeneous effects}  \\
\toprule
$n$ & Method & \multicolumn{3}{c}{ATE} &
\multicolumn{3}{c}{CATE} & \multicolumn{3}{c}{ATE} & \multicolumn{3}{c}{CATE} \\
\midrule \midrule 
 & &rmse& cover&len&rmse&cover&len&rmse&cover&len&rmse&cover&len\\
\cmidrule{3-6} \cmidrule{7-10} \cmidrule{11-14} \cmidrule{15-15}
\multirow{4}{*}{$250$} & BCF   & 0.26  &     0.945 &1.3       & 0.63  &       0.94& 2.5&  0.30 &      0.930 &1.4  &       1.3  &       0.93& 4.5\\
                         & ps-BART &0.54      & 0.780& 1.6    &    1.00 &        0.96 &4.3& 0.56 &      0.805 &1.7  &       1.7   &      0.91 &5.4\\
                         & BART &0.84      & 0.425& 1.5       & 1.20 &        0.90 &4.1 &0.84     &  0.430 &1.6       &  1.8        & 0.87 &5.2\\
                         & BART $(f_0,f_1)$ &1.48     & 0.035 &1.5        &2.42  &       0.80 &6.4 &1.44 &      0.085 &1.6 &        2.6 &        0.83& 7.1 \\
                         & Causal RF   &0.81     &  0.425 &1.5 &       0.84    &     0.70 &2.0&1.10    &   0.305 &1.8      &   1.8       &  0.66 &3.4 \\
                         & LM + HS  &1.77      & 0.015 &1.8 &       2.13         &0.54 &4.4 &1.65     &  0.085 &1.9       &  2.2        & 0.62& 4.8\\
                         \midrule 
\multirow{4}{*}{$500$}    & BCF&     0.20       &0.945& 0.97 &       0.47  &       0.94& 1.9 & 0.23 &      0.910 &0.97 &        1.0  &       0.92& 3.4\\
                         & ps-BART &0.24     &  0.910& 1.07     &   0.62       &  0.99 &3.3& 0.26    &   0.890 &1.06      &   1.1      &   0.95 &4.1 \\
                         & BART &0.31       &0.790 &1.00  &      0.63   &      0.98& 3.0& 0.33&       0.760& 1.00  &       1.1  &       0.94& 3.9\\
                         & BART $(f_0,f_1)$ &1.11     &  0.035& 1.18      & 2.11        & 0.81 &5.8 &1.09     &  0.065& 1.17      &  2.3      &   0.82 &6.2 \\
                         & Causal RF   &0.39      & 0.650 &1.00   &     0.54 &       0.87 &1.7 &0.59  &     0.515& 1.18   &      1.5   &      0.73& 2.8\\
                         & LM + HS  &1.76     &  0.005 &1.34      &  2.19       &  0.40 &3.5 & 1.71    &  0.000 &1.34     &    2.2      &   0.45 &3.7 \\
\bottomrule
\end{tabular*}

\end{table}
\newpage
\subsection{Atlantic causal inference conference data analysis challenges}\label{sec:contest}

The Atlantic Causal Inference Conference (ACIC) has featured a data analysis challenge since 2016. Participants are given a large number of synthetic datasets and invited to submit their estimates of treatment effects along with confidence or credible intervals where available.  Specifically, participants were asked to produce estimates and uncertainty intervals for the sample average treatment effect on the treated, as well as conditional average treatment effects for each unit. Methods were evaluated based on a range of criteria including estimation error and coverage of uncertainty intervals.  The datasets and ground truths are publicly available, so while BCF was not entered into either the 2016 or 2017 competitions we can benchmark its performance against a suite of methods that we did not choose, design, or implement. 

\subsubsection{ACIC 2016 competition} The 2016 contest design, submitted methods, and results are summarized in \cite{dorie2019automated}. Based on an early draft of our manuscript \cite{dorie2019automated} also evaluated a version of BART that included an estimate of the propensity score, which was one of the top methods on bias and RMSE for estimating the sample ATT. BART with the propensity score outperformed BART without the propensity score on bias, RMSE, and coverage for the SATT, and was a leading method overall. 

Therefore, rather than include results for all 30 methods here we simply include BART and ps-BART as leading contenders for estimating heterogeneous treatment effects in this setting. Using the publicly-available competition datasets \citep{aciccomp2016} we implemented two additional methods: BCF and causal random forests as implemented in the R package \texttt{grf} \citep{athey2019generalized}, using 4,000 trees to obtain confidence intervals for conditional average treatment effects and a doubly robust estimator for the SATT (as suggested in the package documentation). 

Table~\ref{tab:acic2016res} collects the results of our methods (ps-BART and BCF) as well as BART and causal random forests. Causal random forests performed notably worse than BART-based methods on every metric. BCF performed best in terms of estimation error for CATE and SATT, as measured by bias and absolute bias. While the differences in the various metrics are relatively small compared to their standard deviation across the 7,700 simulated datasets, nearly all the pairwise differences between BCF and the other methods are statistically significant as measured by a permutation test (Table~\ref{tab:acic2016pval}). The sole exception is the test for a difference in bias between ps-BART and BCF, suggesting the presence of RIC in at least some of the simulated datasets. This is especially notable since the datasets were not intentionally simulated to include targeted selection.  

\cite{dorie2019automated} note that all submitted methods were ``somewhat disappointing'' in inference for the SATT (i.e., few methods had coverage near the nominal rate with reasonably sized intervals). However, ps-BART did relatively well, 88\% coverage of a 95\% credible interval and one of the smallest interval lengths. ps-BART had slightly better coverage than BCF (88\% versus 82\%), with an average interval length that was 45\% larger than BCF. Vanilla BART and BCF had similar coverage rates, but BART's interval length was about 55\% larger than BCF. \cite{dorie2019automated} found that TMLE-based adjustments could improve the coverage of BART-based estimates of the SATT; we expect that similar benefits would accrue using BCF with a TMLE adjustment, but obtaining valid confidence intervals for SATT is not our focus so we did not pursue this further. 

\begin{table}[h!]\footnotesize
\caption{Abbreviated ACIC 2016 contest results. Coverage and average interval length are reported for nominal 95\% uncertainty intervals. Bias and $|$Bias$|$ are average bias and average absolute bias, respectively, over the. PEHE is the average precision in estimating heterogeneous treatment effects (the average root mean squared error of CATE estimates for each unit in a dataset) \citep{hill2011bayesian}.}\label{tab:acic2016res}
\centering
\begin{tabular}{rrrrrrrrr}
  \hline
 & Coverage & Int. Len. & Bias & (SD) & $|$Bias$|$ & (SD) & PEHE & (SD)\\ 
  \hline
BCF & 0.82 & 0.026 & -0.0009 & (0.01) & 0.008 & 0.010 & 0.33 & 0.18 \\ 
  ps-BART & 0.88 & 0.038 & -0.0011 & (0.01) & 0.010 & 0.011 & 0.34 & 0.16 \\ 
  BART & 0.81 & 0.040 & -0.0016 & (0.02) & 0.012 & 0.013 & 0.36 & 0.19 \\ 
  Causal RF & 0.58 & 0.055 & -0.0155 & (0.04) & 0.029 & 0.027 & 0.45 & 0.21 \\ 
   \hline
\end{tabular}
\end{table}

\begin{table}[h!]\footnotesize
\caption{Tests and estimates for differences between BCF and other methods in the ACIC 2016 competition. The p-values are from bootstrapp permutation tests with 100,000 replicates.}\label{tab:acic2016pval}
\centering
\begin{tabular}{rrrrrrr}
  \hline
 & Diff Bias & p & Diff $|$Bias$|$ & p & Diff PEHE & p \\ 
  \hline
ps-BART & -0.00020 & 0.146 & 0.0011 &  $<1e^{-4}$ & 0.010 &  $<1e^{-4}$ \\ 
BART & -0.00070 & $<1e^{-4}$  & 0.0031 &  $<1e^{-4}$ & 0.037 &  $<1e^{-4}$ \\ 
  Causal RF & -0.01453 &  $<1e^{-4}$& 0.0204 &  $<1e^{-4}$ & 0.125 &  $<1e^{-4}$ \\ 
   \hline
\end{tabular}
\end{table}

\subsubsection{ACIC 2017 competition} The ACIC 2017 competition was designed to have average treatment effects that were smaller, with heterogenous treatment effects that were less variable, relative to the 2016 datasets. Arguably, the 2016 competition included many datasets with unrealistically large average treatment effects and similarly unrealistic degrees of heterogeneity\footnote{Across the 2016 competition datasets, the interquartile range of the SATT was 0.57 to 0.79 in standard deviations of $Y$, with a median value of 0.68. The standard deviation of the conditional average treatment effects for the sample units had an interquartile range of 0.24 to 0.93, again in units of standard deviations of $Y$.  A significant fraction of the variability in $Y$ was explained by heterogeneous treatment effects in a large number of the simulated datasets.}. Additionally, the 2017 competition explicitly incorporated targeted selection (unlike the 2016 datasets). The ACIC 2017 competition design and results are summarized completely in \cite{ACIC2017}; here we report selected results for the datasets with independent additive errors. 
\normalsize

\begin{figure}\label{acic2017res}
\includegraphics[width=1.7in,page=1]{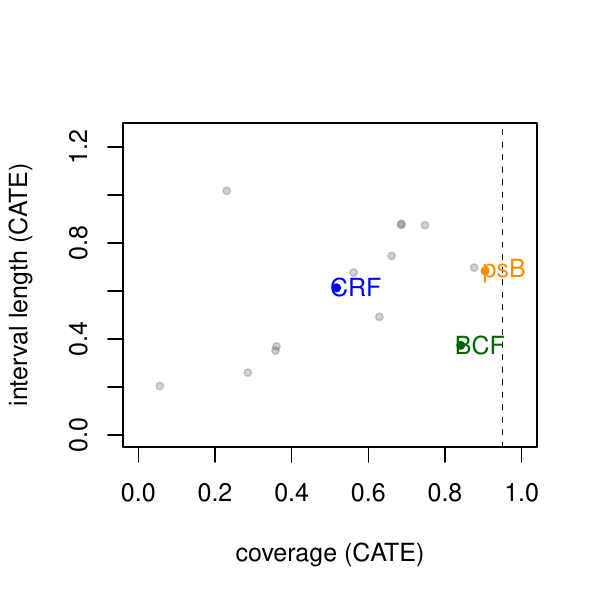}
\includegraphics[width=1.7in,page=2]{figs/test_plots.pdf}
\includegraphics[width=1.7in,page=3]{figs/test_plots.pdf}
\caption{Each data point represents one method. ps-BART (psB, in orange) was submitted by a group independent of the authors based on a draft of this manuscript. TL (purple) is a TMLE-based submission that performed will for estimating SATT, but did not furnish estimates of conditional average treatment effects. BCF (green) and causal random forests (CRF, blue) were not part of the original contest. For descriptions of the other methods refer to \cite{ACIC2017}.}
\end{figure}

Figure \ref{acic2017res} contains the results of the 2017 competition. The patterns here are largely similar to the 2016 competition, despite some stark differences in the generation of synthetic datasets. ps-BART and BCF have the lowest estimation error for CATE and SATE. The closest competitor on estimation error was a TMLE-based approach. We also see that ps-BART edges BCF slightly in terms of coverage once again, although BCF has much shorter intervals. Causal random forests does not perform well, with coverage for SATT and CATE far below the nominal rate.

\section{The effect of smoking on medical expenditures}\label{sec:smoking}
\subsection{Background and data}
As an empirical demonstration of the Bayesian causal forest model, we consider the question of how smoking affects medical expenditures. This question is of interest as it relates to lawsuits against the tobacco industry. The lack of experimental data speaking to this question motivates the reliance on observational data. This question has been studied in several previous papers; see \cite{zeger2000statistical} and references therein. Here, we follow \cite{imai2004causal} in analyzing data extracted from the 1987 National Medical Expenditure Survey (NMES) by \cite{johnson2003disease}.  The NMES records many subject-level covariates and boasts third-party-verified medical expenses. Specifically, our regression includes the following ten patient attributes:

\begin{itemize}
\setlength\itemsep{-0.25em}
\item {\tt age}: age in years at the time of the survey 
\item {\tt smoke\_age}: age in years when the individual started smoking
\item {\tt gender}: male or female
\item {\tt race}: other, black or white
\item {\tt marriage\_status}: married, widowed, divorced, separated, never married
\item {\tt education\_level}: college graduate, some college, high school graduate, other
\item {\tt census\_region}: geographic location, Northeast, Midwest, South, West
\item {\tt poverty\_status}: poor, near poor, low income, middle income, high income
\item {\tt seat\_belt}: does patient regularly use a seat belt when in a car 
\item {\tt years\_quit}: how many years since the individual quit smoking.
\end{itemize}

The response variable is the natural logarithm of annual medical expenditures, which  makes the normality of the errors more plausible. Under this transformation, the treatment effect corresponds to a multiplicative effect on medical expenditure. Following \cite{imai2004causal}, we restrict our analysis to smokers who had non-zero medical expenditure.  
Our treatment variable is an indicator of heavy lifetime smoking, which we define to be greater than 17 {\em pack-years}, the equivalent of 17 years of pack-a-day smoking. See again \cite{imai2004causal} for more discussion of this variable. We scrutinize the overlap assumption and exclude individuals younger than 28 on the grounds that it is improbable for someone that young to have achieved this level of exposure. After making these restrictions, our sample consists of $n = 6,798$ individuals.  


\subsection{Results}
\begin{figure}\label{agetrend}
\includegraphics[width=2in]{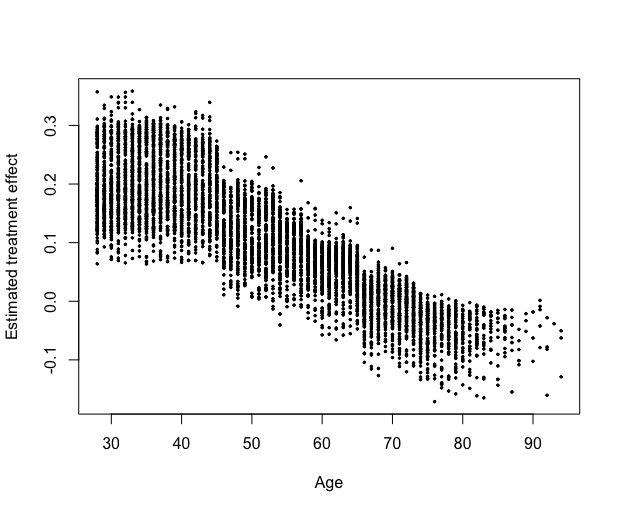}\includegraphics[width=2in]{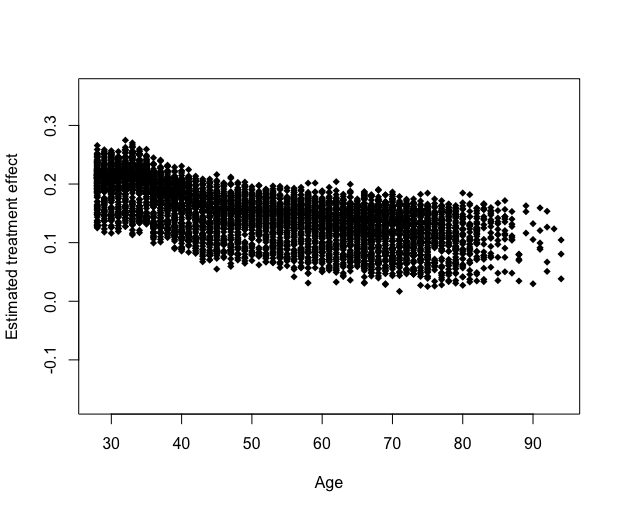}
\caption{Each point depicts the estimated treatment effect for an individual. The BCF model (left panel) detects pronounced heterogeneity moderated by the age variable, whereas BART (right panel) does not.}
\end{figure}

Here, we highlight the differences that arise when analyzing this data using standard BART versus using BCF. First, the estimated expected responses from the two models have correlation of 0.98, so that the two models concur on the nonlinear prediction problem. This suggests that, as was intended, BCF will inherit BART's outstanding predictive capabilities. By contrast, the estimated individual treatment effects are only correlated 0.70. The most notable differences between these CATE estimates is that the BCF estimates exhibit a strong trend in the age variable, as shown in Figure \ref{agetrend}; the BCF estimates suggest that smoking has an pronounced impact on the health expenditures of younger people.  

\begin{figure}\label{RICsmoke}
\includegraphics[width=2in]{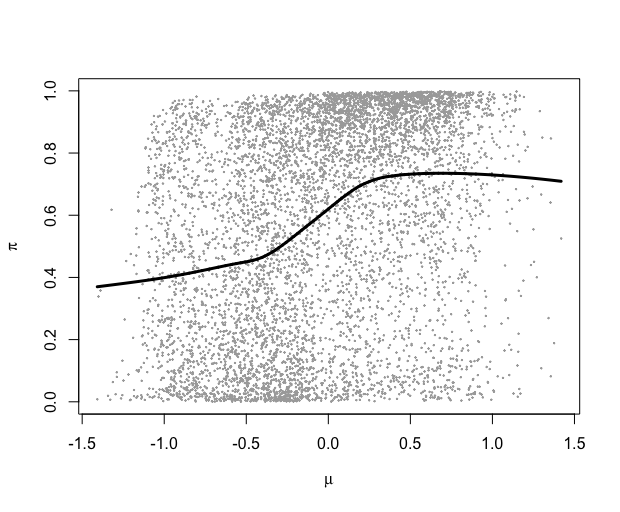}
\caption{Each gray dot depicts the estimated propensity and prognostic scores for an individual. The solid bold line depicts a LOESS trend fit to these points; the monotonicity  is suggestive of targeted selection. Compare to Figures \ref{targeted} and \ref{scatter}. }
\end{figure}

Despite a wider range of values in the CATE estimates (due largely to the inferred trend in the age variable), the ATE estimate of BCF is notably lower than that of BART, the posterior 95\% credible intervals being translated by 0.05, $(0.00, 0.20)$ for BCF vs $(0.05, 0.25)$ for BART.  The higher estimate of BART is possibly a result of RIC. Figure \ref{RICsmoke} shows a LOWESS trend between the estimated propensity and prognostic scores (from BCF); the monotone trend is suggestive of targeted selection (high medical expenses are predictive of heavy smoking) and hints at the possibility of RIC-type inflation of the BART ATE estimate (compare to Figures \ref{targeted} and \ref{scatter}).

\begin{figure}\label{ITE}
\includegraphics[width=2.5in]{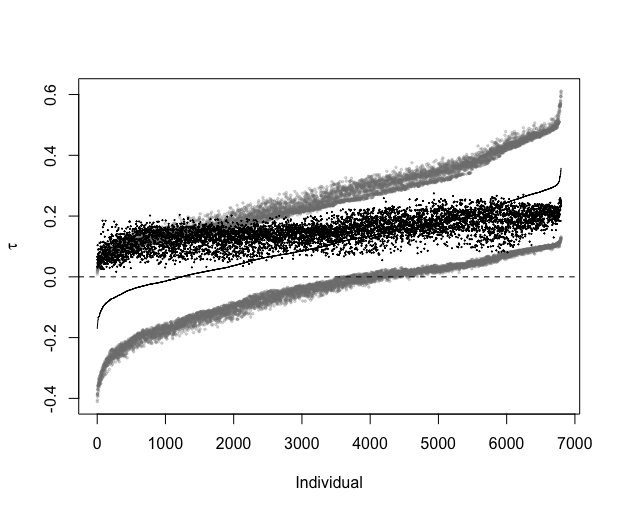}
\caption{Point estimates of individual treatment effects are shown in black. The smooth line depicts the estimates from BCF, which are ordered from smallest to largest. The unordered points represent the corresponding ITE estimates from BART. Note that the BART estimates seem to be higher, on average, than the BCF estimates. The upper and lower gray dots correspond to the posterior 95\% credible interval end points associated with the BCF estimates; most ITE intervals contain zero, especially those with smaller (even negative) point estimates.}
\end{figure}

\begin{figure}\label{rpart}
\includegraphics[width=2.25in]{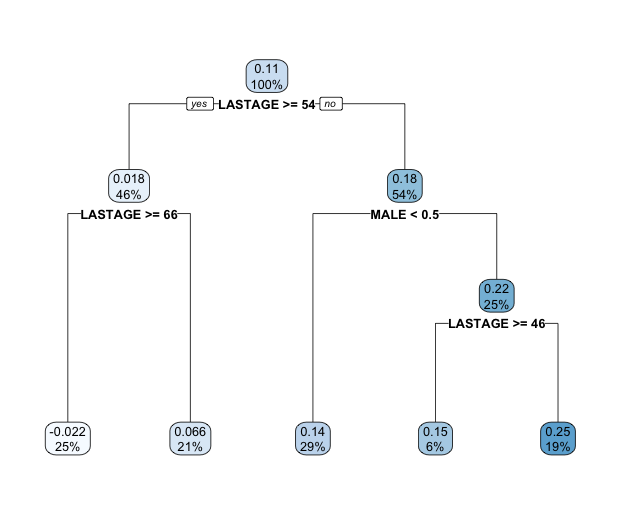}\includegraphics[width=2.25in]{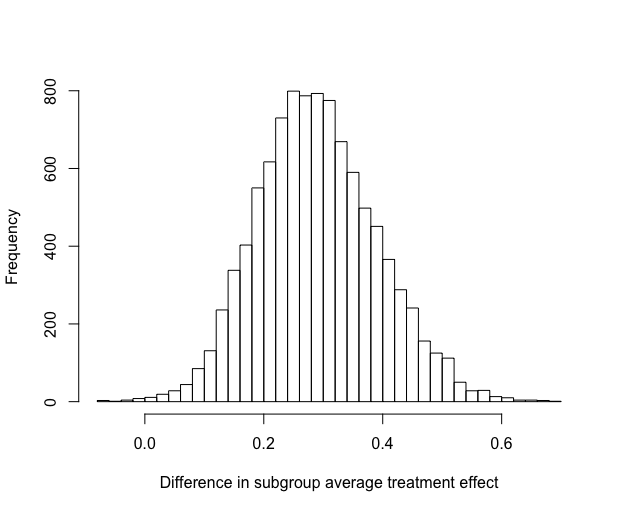}
\caption{Left panel: a summarizing regression tree fit to posterior point estimates of individual treatment effects. The top number in each box is the average subgroup treatment effect in that partition of the population, the lower number shows the percentage of the total sample constituting that subgroup. Age and gender are flagged as important moderating variables. Right panel: based on the tree in the left panel, we consider the difference in treatment effects between men younger than 46 and women older than 66; a posterior histogram of this difference shows that nearly all of the posterior mass is above zero, indicating that these two subgroups are discernibly different, with young men having substantially higher subgroup average treatment effect.}
\end{figure}

Although the vast majority of individual treatment effect estimates are statistically uncertain, as reflected in posterior 95\% credible intervals that contain zero (Figure \ref{ITE}), the evidence for subgroup heterogeneity is relatively strong, as uncovered by the following posterior exploration strategy. First, we grow a parsimonious regression tree to the point estimates of the individual treatment effects (using the {\tt rpart} package in {\tt R}); see the left panel of Figure \ref{rpart}. Then, based on the candidate subgroups revealed by the regression summary tree, we plot a posterior histogram of the difference between any two covariate-defined subgroups. The right panel of Figure \ref{rpart} shows the posterior distribution of the difference between men younger than 46 and women over 66; virtually all of the posterior mass is above zero, suggesting that the treatment effect of heavy smoking is discernibly different for these two groups, with young men having a substantially higher estimated subgroup ATE. This approach, although somewhat informal, is a method of exploring the {\em posterior distribution} and, as such, any resulting summaries are still valid Bayesian inferences. Moreover, such Bayesian ``fit-the-fit'' posterior summarization strategies can be formalized from a decision theoretic perspective \citep{sivaganesan2017subgroup,hahn2015decoupling}; we do not explore this possibility further here.

From the above we conclude that how a model treats the age variable would seem to have an outsized impact on the way that predictive patterns are decomposed into treatment effect estimates based on this data, as age plausibly has prognostic, propensity and moderating roles simultaneously. Although it is difficult to trace the exact mechanism by which it happens, the BART model clearly de-emphasizes the moderating role, whereas the BCF model is designed specifically to capture such trends. Possible explanations for the age heterogeneity could be a mixed additive-multiplicative effect combined with higher baseline expenditures for older individuals or possibly survivor bias (as also mentioned in \cite{imai2004causal}), but further speculation is beyond the scope of this analysis. 

\section{Discussion}\label{discussion}

We conclude by drawing out themes relating the Bayesian causal forest model to earlier work and by explicitly addressing common questions we have received while presenting the work in conferences and seminars. 

\subsection{Zellner priors for non- and semiparametric Bayesian causal inference}

In Section \ref{ric} we showed that the current gold standard in nonparametric Bayesian regression models for causal inference (BART)  is susceptible to regression induced confounding as described by \cite{hahn2016regularization}. The solution we propose is to include an estimate of the propensity score as a covariate in the outcome model. This induces a prior distribution that treats $Z_i$ and $\hat\pi_i$ equitably, discouraging the outcome model from erroneously attributing the effect of confounders to the treatment variable.  Here we justify and collect arguments in favor of this approach. We discuss an argument against, namely that it does not incorporate uncertainty in the propensity score, in a later subsection.

Conditioning on an estimate of the propensity score is readily justified: Because our regression model is conditional on $Z$ and $\X$, it is perfectly legitimate to condition our prior on them as well. 
This approach is widely used in linear regression, the most common example being Zellner's $g$-prior \citep{zellner1986assessing} which parametrizes the prior covariance of a vector of regression coefficients in terms of a plug-in estimate of the predictor variables' covariance matrix. Nodding to this heritage, we propose to call general predictor-dependent priors ``Zellner priors''.

In the Bayesian causal forest model, we specify a prior over $f$ by applying an independent BART prior that includes $\hat{\pi}(\x_i)$ as one of its splitting dimensions. That is, because $\hat{\pi}(\x_i)$ is a fixed function of $\x_i$, $f$ is still a function $f: (\mathcal{X}, \mathcal{Z}) \mapsto \mathbb{R}$; the inclusion of $\hat{\pi}(\x_i)$ among the splitting dimensions does not materially change the support of the prior, but it does alter which functions are deemed more likely. Therefore, although writing $f(\x_i, z_i, \hat{\pi}(\x_i))$ is suggestive of how the prior is implemented in practice, we prefer notation such as
\begin{equation}
\begin{split}
Y_i &= f(\x_i, z_i) + \epsilon_i, \;\;\epsilon_i \iid \N(0, \sigma^2),\\
f &\sim \mbox{BART}(\X, Z, \hat{\pi}),
\end{split}
\end{equation} 
where $\hat{\pi}$ is itself a function of $(\X, Z)$. Viewing BART as a prior in this way highlights the fact that various transformations of the data could be computed beforehand, prior to fitting the data with the default BART priors; the choice of transformations will control the nature of the regularization that is imposed. 
In conventional predictive modeling there is often no particular knowledge of which transformations of the covariates might be helpful. However, in the treatment effect context the propensity score is a natural and, in fact, critical choice.

Finally, some have argued that a committed subjective Bayesian is specifically enjoined from encoding prior dependence on the propensity score in the outcome model based on philosophical considerations \citep{robins1997toward}. We disagree; to the extent that phenomena like targeted selection are plausible, variation in treatment assignment is informative about variation in outcomes under control, and it would be inadvisable for a Bayesian -- committed subjective or otherwise -- to ignore it. 

\subsection{Why not use only the propensity score? vs. Why use the propensity score at all?}
It has long been recognized that regression on the propensity score is a useful dimension reduction tactic \citep{rosenbaum1983central}. For the purpose of estimating average treatment effects, a regression model on the one-dimensional propensity score is sufficient for the task, allowing one to side-step estimating high dimensional nuisance parameters. In our notation, if $\pi$ is assumed known, then one need only infer $f(\pi)$. That said, there are several reasons one should include the control vector $\x_i$ in its entirety (in addition to the propensity score). 

The first reason is pragmatic: If one wants to identify heterogeneous effects, one needs to include any potential effect moderating variables anyway, precluding any dimension reduction at the outset.

Second, if we are to take a conditionally-iid Bayesian regression  approach to inference and we do not in fact believe the response to depend on $\X$ strictly through the propensity score, we simply must include the covariates themselves and model the conditional distribution $p(Y\mid Z, \X)$ (otherwise the error distribution is highly dependent, integrated across $\X$). The justification for making inference about average treatment effects using regression or stratification on the propensity score alone is entirely frequentist; this approach is not without its merits, and we do not intend to argue frequency calibration is not desirable, but a fully Bayesian approach has its own appeal.


Third, if our propensity score model is inadequate (misspecified or otherwise poorly estimated), including the full predictor vector allows for the possibility that the response surface model remains correctly specified. 

The converse question, {\em Why bother with the propensity score if one is doing a high dimensional regression anyway?}, has been answered in the main body of this paper. Incorporating the propensity score (or another balancing score) yields a prior that can more readily adapt to complex patterns of confounding. In fact, in the context of response surface modeling for causal effects, failing to include an estimate of the propensity score (or another balancing score) can lead to additional bias in treatment effect estimates, as shown by the simple, low-dimensional example in Section \ref{ric}.

\subsection{Why not joint response-treatment modeling and what about uncertainty in the propensity score?}

Using a presumptive model for $Z$ to obtain $\hat{\pi}$ invites the suggestion of fitting a joint model for $(Y, Z)$. Indeed, this is the approach taken in \cite{hahn2016regularization} as well as earlier papers, including  \cite{rosenbaum1983central}, \cite{robins1992estimating}, \cite{mccandless2009bayesian}, \cite{wang2012bayesian}, and \cite{zigler2014uncertainty}. While this approach is certainly reasonable, the Zellner prior approach would seem to afford all the same benefits while avoiding the distorted inferences that would result from a joint model if the propensity score model is misspecified \citep{zigler2014uncertainty}. 

One might argue that our Zellner prior approach gives under-dispersed posterior inference in the sense that it fails to account for the fact that $\hat{\pi}$ is simply a point estimate (and perhaps a bad one). However, this objection is somewhat misguided.  First, as discussed elsewhere (e.g. \cite{hill2011bayesian}), inference on individual or subgroup treatment effects follows directly from the conditional distribution $(Y\mid Z, \X)$. To continue our analogy with the more familiar Zellner $g$-prior, to model $(Y\mid Z, \X)$ we are no more obligated to consider uncertainty in $\hat\pi$ than we are to consider uncertainty in $(\X'\X)^{-1}$ when using a $g$-prior for on the coefficients of a linear model. Second, $\hat{\pi}$ appears in the model {\em along with} the full predictor vector $\x$: it is provided as a hint, not as a certainty. This model is at least as capable of estimating a complex response surface as the corresponding model without $\hat{\pi}$, and the cost incurred by the addition of a one additional ``covariate'' can be more than offset by the bias reduction in the estimation of treatment effects.

On the other hand, we readily acknowledge that one might be interested in what inferences would obtain if we used different $\hat\pi$ estimates. One might consider fitting a series of BCF models with different estimates of $\hat\pi$, perhaps from alternative models or other procedures. This is a natural form of sensitivity analysis in light of the fact that the adjustments proposed in this paper only work if $\hat{\pi}$ accurately approximates $\pi$. However, it is worth noting that the available $(\z, \x)$ data speak to this question: a host of empirically proven prediction methods (i.e. neural networks, support vector machines, random forests, boosting, or any ensemble method) can be used to construct candidate $\hat{\pi}$ and cross-validation may be used to gauge their accuracy. Only if a ``tie'' in generalization error (predicting $Z$) is encountered must one turn to sensitivity analysis.

\subsection{Connections to doubly robust estimation}
Our combination of propensity score estimation and outcome modeling is superficially reminiscent of doubly robust estimation \citep{bang2005doubly}, where propensity score and outcome regression models are combined to yield consistent estimates of finite dimensional treatment effects, provided at least one model is correctly specified.  We do not claim our approach is doubly robust, however, and in all of our examples above we use the natural Bayesian estimates of (conditional) average treatment effects rather than doubly robust versions.  Empirically these seem to have good frequency properties.  The motivation behind our approach (the parameterization and including an estimate of the propensity score) is in fact quite different than that behind doubly robust estimation, as we are not focused on consistency under partial misspecification, nor obtaining parametric rates of convergence for the ATE/ATT, but rather on regularizing in a way that avoids RIC. To our knowledge, none of the literature on doubly robust estimators explicitly considers bias-variance trade-off ideas in this way. 

Although it was not our focus here we do expect that BCF would perform well in the context of doubly robust estimation. For example, a TMLE-based approach using SuperLearner with BART as a component of the ensemble was a top performer in the 2017 ACIC contest. BCF and ps-BART generally improve on vanilla BART, and should be useful in that context. As another example, \cite{nie2017quasi} showed that using ps-BART (motivated by an early draft of this paper) as a component of a stacked estimator of heterogeneous treatment effects fitted using the $R-$learner yielded improved performance over the individual heterogeneous treatment effect estimators.  We hope that researchers will continue to see the promise of BCF and related methods as components of estimators derived from frequentist considerations.

\subsection{The role of theory versus simulations in methodological comparisons.}
Recent theory on posterior consistency and rates of posterior concentration for Bayesian tree models in prediction contexts \citep{linero2018bayesian,rockova2017posterior,rockova2019theory} should apply to the BCF parametrization with some adaptation. However, the existing results require significant modifications to the BART prior that may make them unreliable guides to practical performance. Likewise, recent results demonstrate the consistency of recursive approximations to single-tree Bayesian regression models \citep{he2019stochastic} in the setting of generalized additive models and these results are possibly applicable to BCF type parametrizations. 

Despite only nascent theory  --- and none speaking to frequentist coverage of Bayesian credible intervals --- BCF should be of interest to anyone seeking reliable estimates of heterogeneous treatment effects, for two reasons. The first is that many of the existing approaches to fusing machine learning and causal inference make use of first-stage regression estimates for which no dedicated theory is strictly necessary, for instance \cite{kunzel2019metalearners} and \cite{nie2017quasi}. In this context, BCF can be regarded as another supervised learning algorithm, alongside neural networks, support vector machines, random ,forests, etc. and would be of special interest insofar as it obtains better first-stage estimates than these other methods. 

The second, and more important reason, that BCF is an important development is its performance in simulation studies by us and others. Unlike many simulation studies, designed with the goal of showcasing a method's strengths, our simulation studies were designed prior to the model's development, with an eye towards realism. Specifically, our simulation protocol was created specifically to correct perceived weaknesses in previous synthetic data sets in the causal machine learning literature:  no or very weak confounding, implausibly large treatment effects, and unrealistically large variation in treatment effects (including sign variation). By contrast, our data generating processes reflect our assumptions about real data for which heterogeneous treatment effects are commonly sought: strong confounding, and modest treatment effects and treatment effect heterogeneity (relative to the magnitude of unmeasured sources of variation). It was these considerations that led us to the concept of targeted selection (Section~\ref{sec:targetedselection}), for example.

By utilizing realistic, rather than convenient or favorable, data generating processes, our simulations are a principled approach to assessing the finite sample operating characteristics of various methods. Not only did this process reassure us that BCF has good frequentist properties in the regimes we examined, but it also alerted us to what cold comfort asymptotic theory can be in actual finite samples, as methods with available theory did not perform as well as the theory would suggest. Finally, we would also note that other carefully designed simulation studies reach similar conclusions (e.g. ACIC 2016, described above, and \cite{wendling2018comparing,mcconnell2019estimating}).

\bibliographystyle{ba}
\bibliography{treateffectJCGS}

\end{document}